Abstract: This study investigates the impact of a fiscal policy spending shock on the economy of the Visegrad 4 countries. The impact is estimated with an SVAR model, and the calculations are based on 84 quarterly observations (1999Q1-2019Q4). The results suggest that fiscal expansion has a larger than usual impact in the V4 countries (except for Slovakia): the estimated long-term (5-year) cumulative spending multipliers are 0.81 for Czechia, 1.14 for Hungary, and 1.76 for Poland (the Slovakian multiplier has a value of -0.18, but it is not significant). The discussion section also connects higher spending multipliers with a higher share of VAT revenues, a higher debt ratio, higher foreign debt, and lower openness.

Keywords: fiscal policy, government spending multiplier, SVAR, Visegrad countries

Introduction

Slowing growth has become one of the major economic policy concerns of the 2020s. Inflationary pressure, the energy crisis and rising interest rates have led to a global slowdown. The IMF expects a 3 percent growth in 2023 (compared to 3.5% in 2022; IMF 2023b), but in Central and Easter Europe (CEE), the decline will probably be much sharper: the growth of the Polish economy is forecasted to be 0.3 percent in 2023, compared to 5.4 percent in 2022 (IMF 2023a). Monetary policy has become tighter compared to the 2010s, and interest rates are particularly high in the CEE region: while the ECB rate is at 3%, the Polish policy rate is 6.75%, the Czech policy rate is 7%, and the Hungarian policy rate is 13%. Rising rates make fiscal expansion more expensive, but fiscal policy may well be one of the very few options for governments if they want to generate growth. According to the IMF, government support paid to households and firms worth 1.2 percent of the EU GDP was the reason why European growth was more resilient than expected in 2022 (IMF 2023a).

The aim of this paper is to investigate the impact of fiscal policy in the Visegrad countries (V4; Czechia, Hungary, Poland, and Slovakia) and to identify factors that may influence its efficiency. We chose V4 for multiple reasons. These countries are expected to be hit particularly hard by the global economic slowdown, which will draw increased attention to fiscal stimulus. The geopolitical environment of the V4 is very similar; they are at a similar level of development, while there is a significant variation among the countries in factors such as openness, exchange rate regime, tax regime, or monetary policy that seem to have an impact on the efficiency of fiscal shocks (Ramey 2019). Empirical research on the topic is also very scarce among the V4.

Fiscal policy impact is usually assessed through the concept of fiscal multipliers. The concept was originally developed in the Keynesian demand-side framework, but the neoclassical model focusing on supply-side channels also identifies spending and tax multipliers. The New Keynesian school merges the two different concepts using dynamic stochastic general equilibrium (DSGE) models (Ramey 2019). Empirical studies on fiscal multipliers were based on two main approaches: a model-based approach using simulations (DSGE models) and an econometric approach (mostly applying vector autoregressive (VAR) models). VAR models also differ according to their specifications (e.g., set of endogenous variables or deterministic terms) and by the method chosen to identify fiscal shocks (Caldara–Kamps 2008). Caldara and Kamps (2008) list four different identification approaches: the recursive approach (Fatas–Mihov 2001); the structural VAR approach with an external elasticity coefficient (Blanchard–Perotti 2002b); the sign-restrictions approach (Mountford–Uhlig 2009); and the event-study (narrative) approach (Ramey–Shapiro 1998).

This study adopts the structural VAR (SVAR) approach originally developed by Blanchard and Perotti (2002). An independent SVAR model is estimated for each of the V4 countries, in which the government expenditure variable is calculated by subtracting subsidies from the total government expenditure, and value added tax (VAT) revenues are used as a proxy for government revenues. Our models are based on quarterly data for the period of 1999Q1-2019Q4.

We make contributions to the field of fiscal multipliers in three areas. We present an estimation for the V4 countries by applying the Cholesky decomposition, and we detect significant long-run expenditure multipliers surpassing unity in two countries (Hungary and Poland). By using VAT revenues as a proxy for government revenues, our results could support the proposition that the use of nondistortionary taxes could be associated with higher multipliers. Finally, we offer a discussion of potential economic variables that may also have an impact on the size of the multiplier and thus on the efficiency of fiscal policy interventions.

The rest of this study is organised as follows. Section 2 provides a literature review on the fiscal multiplier and on the macroeconomic conditions that were found to impact its size. In Section 3, we describe the data and the model used for estimating the expenditure multiplier in the V4 countries. Section 4 presents and discusses the main results of our analysis. The conclusion section summarises our findings and points to some implications and limitations.

Literature review

Blanchard and Perotti (2002) investigated US fiscal policy and GDP data for the postwar period spanning over 50 years (1947Q1 to 1997Q4). They calculated multipliers as the ratio of the output response at a particular horizon to the impact effect of the shock on government spending. Depending on what horizon was chosen (it could be the period with the peak output response or any number of periods/years following the shock), the spending multiplier had a peak value no larger than 1. This triggered much research in the field, as stylised facts derived from the Keynesian framework would have suggested larger than 1 multiplier in the case of spending shocks. Similar SVAR estimations made on a sample of OECD countries (Australia, Canada, the UK, USA, and West Germany) over 40 years (1960-2000) (Perotti, 2005) or on Euro area countries for the period 1981Q1 to 2007Q4 (Burriel et al. 2010) found vaguely similar spending multipliers.

The above studies point out that the multiplier value changes over time, and in some periods, it is much higher than the overall average. This variance in the multiplier suggests that the context of the economic environment has a large impact on its value. Estimations may also be influenced by the model specifics (e.g., the method of identifying shocks, the lag built into the system, the included variables). Caldara and Kamps (2008) test US data from 1955-2006 and find that if they control for differences in model specifications, all SVAR approaches yield very similar government spending shocks; the multiplier is typically above 1 and peaks at a value of approximately 2. Applying a modified (regime-switching) SVAR model (Auerbach–Gorodnichenko 2012) also estimates US spending multipliers (sample period 1947Q1-2008Q4). They find values between 0 and 0.5 in expansion periods and between 1 and 1.5 in recessions.

Regarding the geographical context, spending shocks seem to be within a 0.5-1.5 range. A Portuguese estimation based on the Blanchard-Perotti method over the 1995-2017 period finds significant spending multipliers higher than 1 (Bova–Klyviene 2020). An augmented Blanchard-Perotti SVAR yields spending multipliers below 1 for Turkey for the 2002-2016 period (Şen–Kaya 2020). A study from Danish researchers on the same SVAR approach and a longer time series (1983-2017) finds a spending multiplier of approximately 0.7 for Denmark (Kronborg 2021). Another SVAR estimate on Italian data for the 1995-2019 period yields a below 1 spending multiplier (Deleidi 2022). (van Rensburg et al. 2022) evaluate the fiscal policy of South Africa; they find that the fiscal expenditure multiplier declined from 1.5 in 2010 to near 0 in 2019.

Only a limited number of studies are available on the V4 countries. One such attempt investigated a rather short (2000Q1-2012Q4) period with an SVAR model similar to Auerbach and Gorodnichenko's (2012). The authors find a combined V4 spending multiplier that is higher than those detected in the US or in the Eurozone (Baranowski et al. 2016). (Mirdala–Kameník 2017) focus on Czechia, Hungary and Slovakia (period: 1995-2015). They calculate the impact of a fiscal shock with a TVAR model and find positive and significant spending multipliers (the highest for Slovakia and the lowest for Czechia). A study focusing on Hungary and Slovakia found positive spending multipliers in both countries (Kameník et al. 2018). The most recent study compares Czechia, Hungary and Poland with a Blanchard-Perotti SVAR model in the 2002-2018 period (Szymańska 2019). Szymanska calculates positive spending multipliers (the lowest for Czechia and the highest for Poland); the Polish peak multiplier is significantly higher than 1 (and in fact is very close to 2).

The regime-switching SVAR approach already mentioned in this review was introduced to distinguish between the low and high points of the economic cycle. This is one element of the economic context that has an influence on fiscal policy efficiency, but there are several other aspects as well discussed in the literature. We will introduce six of these in this review: economic cycle (recession/expansion); method of financing (quantitative easing/debt); fiscal space and other aspects of government debt; level of interest rates; trade openness; and tax regime (distortionary/nondistortionary taxes).

In regard to the effect of the economic cycle, it was found by many studies that government expenditure has a much larger impact on output in recession than in expansion periods (it can be zero or even negative in this latter case) (Auerbach–Gorodnichenko 2012, Baranowski et al. 2016). Bentour (2022) examines 18 OECD countries (the V4 are not included) between 1966 and 2019 with an SVAR model that takes the debt level into consideration and finds that fiscal expansion has a more dynamic impact in recession periods. Contrary to these results, (Boitani et al. 2020) do not find evidence for larger multipliers in recession periods on a sample of 10 Eurozone countries between 1992 and 2015. They use a regime-switching SVAR and find spending multipliers higher than 1.

Macroeconomic expansion may be triggered by quantitative easing or by a fiscal stimulus in the form of government spending. Both methods have their costs according to stylised facts: while monetary expansion may trigger inflation, an increase in public debt results in higher future taxes (Kocziszky et al. 2018). Using New Keynesian DSGE models, researchers typically find that a money-financed fiscal stimulus has much larger multipliers than a debt-financed fiscal stimulus (Galí 2020, Mao et al. 2023). However, money financing does not seem to be an option in the high-inflation environment of the 2020s. The institutional framework can also have an impact on the effectiveness of fiscal expansions. Higher levels of corruption (Rontos et al. 2023) or political budget cycles (Petrakos et al. 2021) in the V4 countries may lower efficiency.

The fiscal space of an economy and the government debt to GDP ratio may influence fiscal policy through several channels. According to the Ricardian equivalence, a fiscal stimulus could prompt larger future taxes, especially if the debt ratio is high. Government borrowing can also drive private investment out by raising the interest rate. A high public debt/GDP ratio (or smaller fiscal space) could therefore reduce the effectiveness of fiscal policy, but there are also some counterbalancing forces. At a zero lower bound, government borrowing may not increase the interest rate. Similarly, if the government has access to foreign savings, this could also limit the growth of internal interest rates. A higher debt/GDP ratio usually means a lower propensity to save, and the Keynesian framework suggests that ceteris paribus, the lower the propensity to save, the higher the multiplier. Estimating on an OECD sample, (Auerbach–Gorodnichenko 2017) find that fiscal stimulus can even improve fiscal sustainability (so the impact on GDP is larger than that on the debt level). On a 10-country Eurozone sample, it is found that expenditure multipliers are higher in high debt economies (Boitani et al. 2020). A similar effect is found using a New Keynesian model (Aloui–Eyquem 2019). Other SVAR studies, however, conclude that lower fiscal space can be associated with lower spending multipliers (Deskar-Škrbić–Šimović 2017, Huidrom et al. 2020, Kamenik–Stiblarova 2021).

SVAR estimations have shown that the fiscal space impact is moderated by the effect of foreign financing. The size of fiscal multipliers increases as the share of government debt held by foreigners increases or if the expansion is financed from foreign sources (Broner et al. 2022, Farhi–Werning 2016). FDI flows also play an important role in the size of the fiscal multiplier (Wierzbowska–Shibamoto 2018).

The level of policy rates can influence the multiplier size because at near zero interest rates (zero or effective lower bound), it is much less likely that an increase in government debt raises the market rates. This was confirmed by a panel SVAR analysis conducted on Eurozone economies (Amendola et al. 2020). The calculations suggest that cumulated multipliers are much higher at periods of effective zero bound than at normal times (1.6-2.9 and 0.3-1.4). No such influence was detected in US historical data (Ramey–Zubairy 2018). Zero lower bound discussions are less important in times of interest rate hikes.

Trade influences the efficiency of fiscal stimulus through imports. In the Keynesian framework, the marginal propensity to import decreases the multiplier because the extra government expenditure may be spent on foreign goods. (Ilzetzki et al. 2013) build a Blanchard-Perotti-type SVAR model for 44 countries and find that countries with a high trade/GDP ratio have negative multipliers, while the multiplier of countries with less trade is significantly higher than zero. The same study also finds that the fiscal multiplier is higher in countries with fixed exchange rates than in countries with flexible exchange rate regimes. Other studies find a similar connection between openness and fiscal multiplier size (Deskar-Škrbić–Šimović 2017, Riguzzi–Wegmueller 2017).

Finally, tax policies can also be growth-enhancing, leading to higher multipliers. A meta-analysis of 49 studies conducted on OECD samples concludes that revenues coming from distortionary taxes spent on unproductive expenditures are significantly associated with lower economic growth, while nondistortionary tax revenues used to fund productive expenditures are significantly associated with higher economic growth (Alinaghi–Reed 2021). Consumption taxes and VAT are considered to be nondistortionary (Kneller et al. 1999); therefore, we should expect higher multipliers in countries that gain a larger proportion of government revenue from such taxes.

Studies focusing on the Visegrad countries have found similar fiscal influencing factors. Simionescu et al. (2017) found that FDI promoted economic growth in Czechia, Hungary, and Poland (but not in Slovakia). Győrffy (2022), on the other hand, points out that strategies solely focusing on foreign investments can lead to a middle-income trap in Central and Eastern Europe. Sávai (2019), while focusing on government debt, finds a connection between the GDP growth rate and general government debt.

Data and method

Data on government expenditure, government revenues and output for all the countries are from Eurostat; short-term interest rates come from the OECD database; the source of GDP, inflation and short-term interest rate statistics for the USA is also the OECD database. We use the VAT as a proxy for government revenues. This series is not seasonally adjusted; thus, we apply the TRAMO-SEATS procedure to obtain the seasonally adjusted series. Data on government expenditure and output are provided by Eurostat with seasonal adjustment. We subtract the subsidies from the government expenditure because they are the most cyclical component of spending, thus helping in the identification. We also apply the TRAMO-SEATS on the time series of subsidies because it is not available in seasonal adjusted form. All the variables are transformed into real by using the consumer price index of each country. Data are downloaded from Eurostat - Quarterly nonfinancial accounts for general government, and OECD - Short-term interest rates for Visegrad countries: Poland, Czech Republic, Slovakia, and Hungary.

Our model uses 84 quarterly observations over the period of 1999Q1-2019Q4. Although later data are also available, we decided to disregard the pandemic years, as the empirical approach we use does not handle the major break in the time series well. Dropping periods with extreme fluctuations is a solution that is suggested in the literature (Lenza–Primiceri 2022). The endogenous variables of the analysis are total general government expenditure, subsidies, VAT, GDP, and short-term interest rate. We also added the following exogenous factors (as control variables): USA GDP, Inflation rate, and Short-term interest rate.

To study the effects of fiscal policy on output and estimate the government expenditure multipliers, we use the SVAR approach. We estimate a model for each country. The variables that enter the VAR are government expenditure, $G_t$, government revenue, $T_t$, output, $Y_t$, and short-term interest rate $i_t$. They are transformed as in Hall (2009) and Mumtaz and Sunder-Plassman (2021), namely, the first difference of each variable is divided by the lag of output: (i) $G_t = (\tilde{G}_t - \tilde{G}_{t-1})/\tilde{Y}_{t-1}$; (iii) $T_t = (\tilde{T}_t - \tilde{T}_{t-1})/\tilde{Y}_{t-1}$; (iii) $Y_t = (\tilde{Y}_t - \tilde{Y}_{t-1})/\tilde{Y}_{t-1}$, where the tilde indicates the level of the variables. This ensures that all the variables are expressed in the same unit, helping in the computation of the multipliers (more on this later). The reduced form VAR(4) is as follows:

$$X_t = c + \Gamma_1 X_{t-1} + \Gamma_2 X_{t-2} + \Gamma_3 X_{t-3} + \Gamma_4 X_{t-4} + u_t \qquad (1)$$

where $X_t = [G_t, T_t, Y_t, i_t]$, $c$ is a vector of constants, $\Gamma_i$ is a $4 \times 4$ matrix containing the coefficients of the i-th lag, and $u_t = [u_t^G, u_t^T, u_t^Y, u_t^i]$ is the vector of the innovations. The structural form is represented as in the so-called B-model (Lütkepohl 2005), where the innovations are linked to the structural shocks with a matrix B that contains the coefficients measuring the impact responses of the endogenous variables to the structural shocks $\varepsilon_t$:

$$u_t = B\varepsilon_t \qquad (2)$$

To identify this matrix, we assume a recursive scheme taking some ideas from the Blanchard and Perotti (2002) approach relative to the lags in the implementation of fiscal policy and impose short-run restrictions. We assume that discretionary fiscal policy does not respond contemporaneously to an output shock within a quarter because of lags in the implementation of fiscal policy and because policy-makers do not have immediate access to GDP information due to delays in releasing the data. Therefore, we identify matrix B, applying the Cholesky decomposition of the reduced-form covariance matrix, which implies the following restrictions:

$$\begin{bmatrix} u_t^G \\ u_t^T \\ u_t^Y \\ u_t^i \end{bmatrix} = \begin{bmatrix} - & 0 & 0 & 0 \\ - & - & 0 & 0 \\ - & - & - & 0 \\ - & - & - & - \end{bmatrix} \begin{bmatrix} \varepsilon_t^G \\ \varepsilon_t^T \\ \varepsilon_t^Y \\ \varepsilon_t^i \end{bmatrix} \qquad (3)$$

We order government expenditure first; thus, this means that government expenditure does not react to shocks hitting GDP within a quarter. Then, we are interested in the dynamic response of the variables to a government spending shock; therefore, we compute the impulse response functions (IRFs), which measure the response of the endogenous variables at different horizons after the shock:

$$\Delta X_{t+h} = F^h B \qquad (4)$$

where $F$ is the companion form matrix. Furthermore, we aim to estimate the multipliers. The transformation of the variables that we adopt, as shown above, ensures that all the variables are expressed in the same unit and that the multipliers can be estimated by dividing the cumulative IRF of GDP to a government spending shock by the cumulative IRF of government expenditure to a government spending shock:

$$m_h = \frac{\sum_{h=0}^{20} \Delta Y_{t+j}}{\sum_{h=0}^{20} \Delta G_{t+j}} \qquad (5)$$

The inference is based on the bootstrap proposed by Runkle (2002), namely, after estimating the VAR, we use the estimated coefficients and residuals to simulate data by sampling with replacement of the residuals. We perform 1000 replications, and at each iteration, we store the IRFs and the multipliers. Then, we compute 68% and 90% confidence intervals, taking the 16th and 84th quantiles and the 5th and 95th quantiles of the simulated distribution of the IRFs and multipliers (the 68% confidence interval is commonly used in the literature, e.g., Deleidi et al. 2021, Matarrese–Frangiamore 2023). In our figures, we represent the response of the variables in levels by cumulating the IRFs.

Results and discussion

The 4 SVAR models run for the V4 countries gave us 4 IRFs (spending, revenues, output and short-term interest rate) for each country and the cumulative multipliers calculated as in Equation 5. The estimated IRFs assume a government spending shock of one standard deviation, with the shock's dynamics subject to change over the course of 20 quarters. An exogenous rise in government expenditure is typically followed by a persistent dynamic, suggesting that an initial government spending shock may build up over time. The IRF depicts the dynamic effect at a certain horizon of the response variable after an initial shock, while the cumulative multiplier depicts the reaction of output per unit of government expenditure. Our technique estimates cumulative multipliers by dividing the total change in output by the total change in the public expenditure under consideration; namely, we divide the cumulative IRF of output to a government spending shock by the cumulative IRF of government expenditure to a government spending shock.

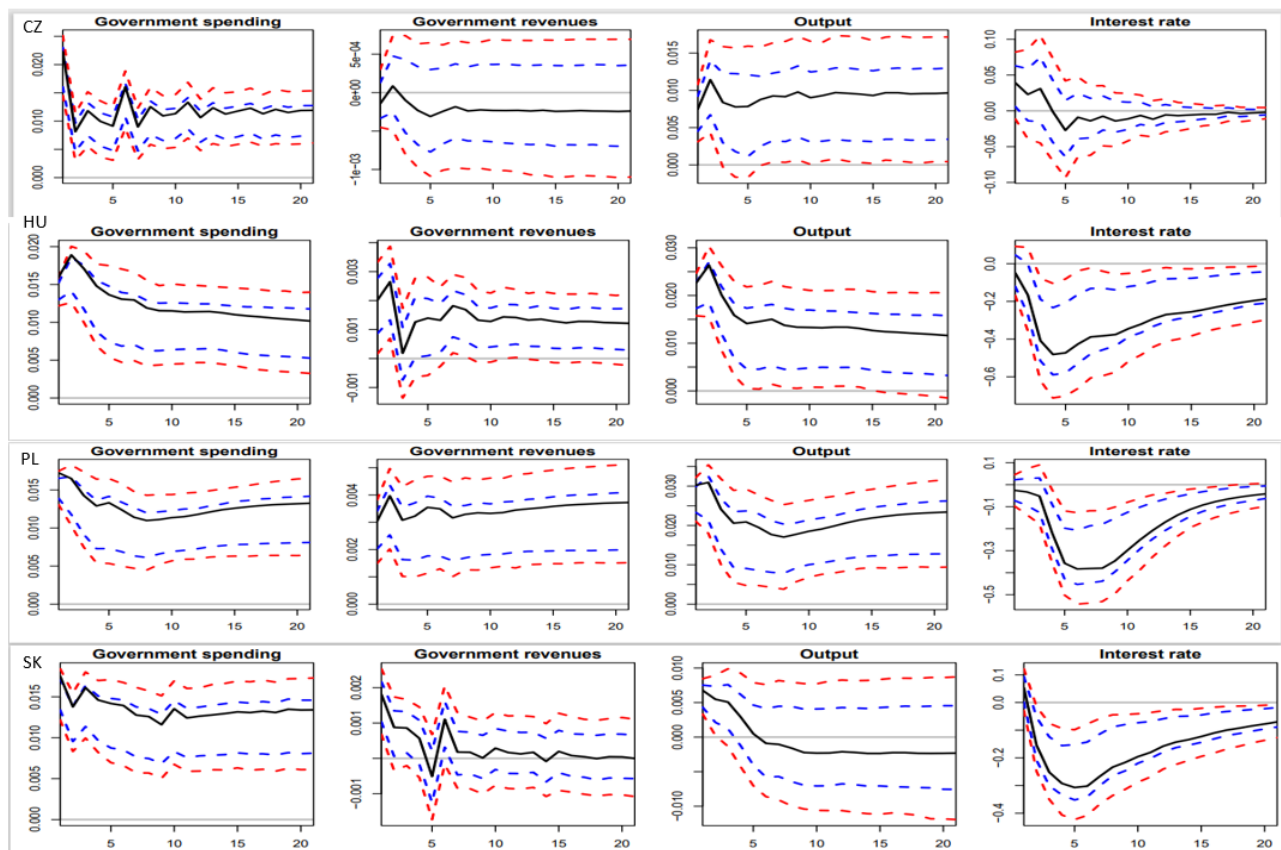

Figure 1. Impulse response functions in the V4 countries

Description: The order of the countries is – Czechia on top, then Hungary, Poland, and Slovakia at the bottom. The black lines represent the estimated IRF values; the blue dashed line shows the 68% interval bands (16th and 84th quantiles), and the red dashed line shows the 90% interval bands.

Source: own calculations

Table 1. Cumulative government spending multipliers for the V4 countries

| Horizont | Czechia | Hungary | Poland | Slovakia |
|---|---|---|---|---|
| Q1 | 0.335* | 1.420** | 1.753** | 0.386** |
| Q2 | 1.405* | 1.396** | 1.876** | 0.398** |
| Q3 | 0.709** | 1.177** | 1.692** | 0.315* |
| Q4 | 0.780** | 1.072** | 1.595** | 0.199* |
| Q5 | 0.859** | 1.036** | 1.569** | 0.034 |
| Q6 | 0.546** | 1.114** | 1.572** | -0.064 |
| Q7 | 1.030* | 1.160** | 1.551** | -0.082 |
| Q8 | 0.730* | 1.156** | 1.552** | -0.126 |
| Q9 | 0.894* | 1.159** | 1.598** | -0.192 |
| Q10 | 0.796* | 1.156** | 1.631** | -0.171 |
| Q11 | 0.699* | 1.164** | 1.660** | -0.185 |
| Q12 | 0.912* | 1.173** | 1.687** | -0.167 |
| Q13 | 0.777* | 1.170** | 1.709** | -0.173 |
| Q14 | 0.843* | 1.162** | 1.725** | -0.182 |
| Q15 | 0.789* | 1.150* | 1.737** | -0.180 |
| Q16 | 0.787* | 1.145* | 1.746** | -0.172 |
| Q17 | 0.856* | 1.146* | 1.752** | -0.169 |
| Q18 | 0.788* | 1.147* | 1.757** | -0.179 |
| Q19 | 0.830* | 1.147* | 1.761** | -0.175 |
| Q20 | 0.806* | 1.144* | 1.764** | -0.175 |

Description: Values with a ** represent statistically significant results at the 90% confidence level; * shows statistically significant results at the 68% confidence level.

Source: own calculations

Table 1 indicates that the estimated impact of a fiscal shock on GDP is most intense in Poland and Hungary. Over the first 4 quarters, the calculated spending multiplier is significant in all countries, but the spread is surprisingly large: while the Hungarian and especially the Polish multipliers are steadily above 1, the Czech one fluctuates a lot but stabilises below 1, and the multiplier calculated for Slovakia does not even reach 0.5 and is not even significant from the fifth quarter. Figure 2 is a graphical representation of the calculated V4 multipliers. It not only shows the estimated multiplier value (black line) but also the 68% and 90% confidence intervals (in the form of the red and blue dashed lines).

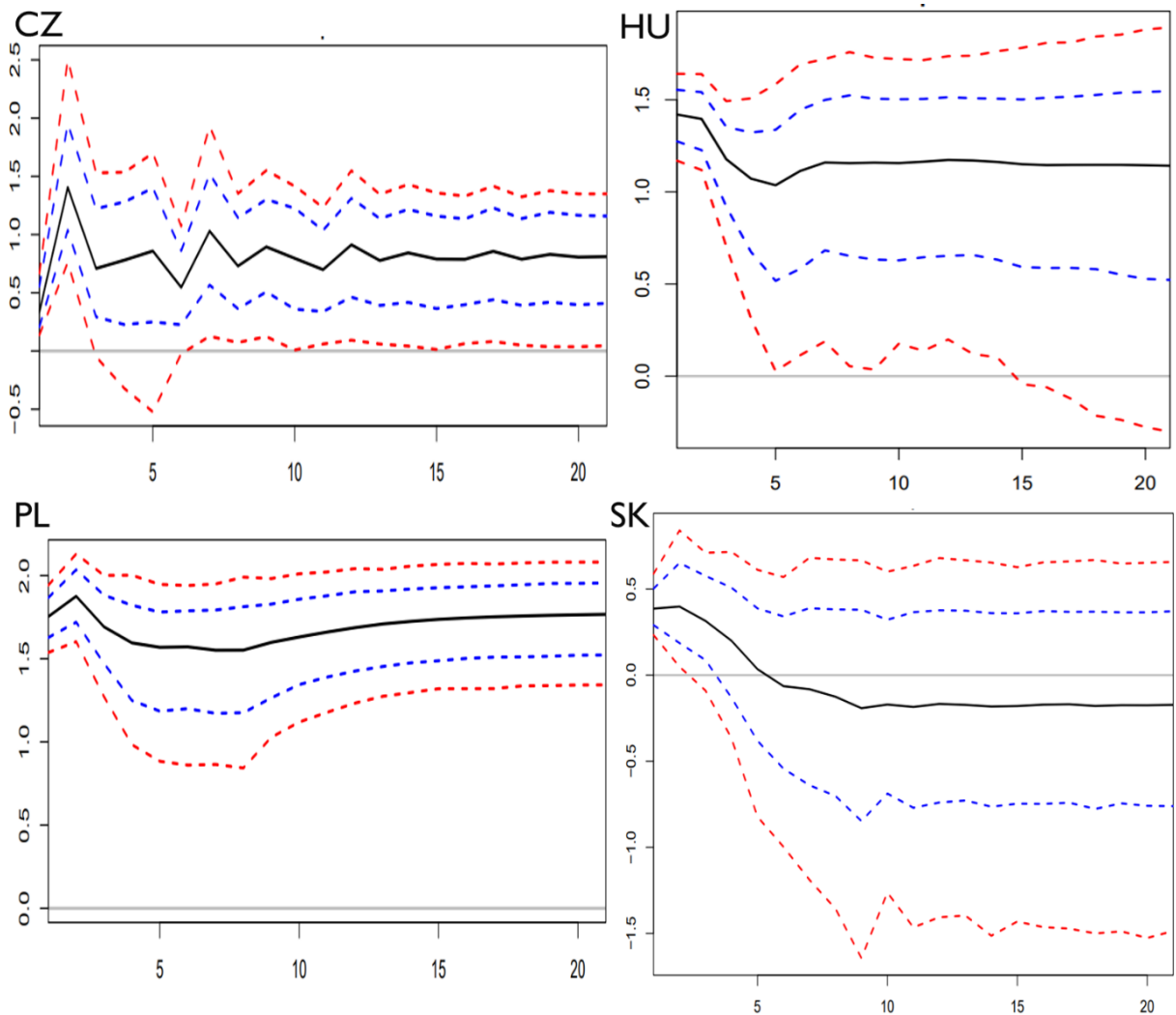

Figure 2. Spending multipliers in the V4 countries with 90% and 68% confidence intervals

Description: The black lines represent the estimated multiplier values; the blue dashed line shows the 68% interval bands (16th and 84th quantiles), and the red dashed line shows the 90% interval bands.

Source: own calculations

Some studies focus on the peak value of the multiplier. In our models, the multiplier peaks in the second quarter in three countries (Czechia, Poland, and Slovakia) and in the first quarter in Hungary. The use of the peak value could be somewhat misleading, but the cumulative multipliers even over longer horizons (e.g., at the end of the 5th year when our estimations end) are only approximately 0.2-0.6 lower than at their peak.

Our estimated spending multipliers, especially in the case of Hungary and Poland, are considerably higher than 1 and higher than what was traditionally found in more advanced economies. Baranowski et al., (2016) investigate a much shorter period and combine the data of the V4 countries, but they also find much higher spending multipliers in this region. According to their results, the spending multiplier in the V4 countries was 1.17 in the 2000-2009 (expansion) period and 3.85 in the 2009-2012 (recession) period. Mirdala and Kameník (2017) find lower multipliers for Czechia (0.3 or smaller) and Hungary (0.1-0.6) but a slightly higher one for Slovakia (0.5-0.7). The most recent study for the region was conducted by Szymańska (2019). She finds fairly similar spending multipliers: 0.4-1.2 for Czechia, 0.5-1 for Hungary, and 1.4-1.8 for Poland.

Our results seem to confirm that fiscal expansion can have a larger impact in the CEE-V4 region. We had 84 quarterly observations available, which is the longest time series used for IRF and multiplier estimation in the region. Another difference in our approach compared to other V4 attempts is that we subtracted the subsidies from the total government expenditure to boost the identification process. This could explain why we found higher or slightly higher spending multipliers in Czechia, Hungary, and Poland than some of the previous studies. Furthermore, our data were expressed in terms of euros, meaning that they included the impact of exchange rate changes (earlier studies found a connection between exchange rate regimes and the effectiveness of fiscal stimulus).

Table 2. Average indicator values for the V4 countries

| Indicator | Czechia | Hungary | Poland | Slovakia |
|---|---|---|---|---|
| Government debt (% of GDP) *OECD, 1995-2019* | 35.6% | 79.2% | 57.5% | 51.3% |
| Interest rates (EMU conv. crit.) *Eurostat, 2001-2019* | 3.2% | 6.1% | 5.1% | 3.6% |
| Share of imports (% of GDP) *OECD, 1995-2019* | 58.8% | 68.9% | 38.7% | 75% |
| FDI stock (% of GDP) *OECD, 2005-2019* | 59.5% | 66.5% | 37.5% | 57.1% |
| Net external debt (% of GDP) *Eurostat, 2004-2019* | -8.2% | 36.6% | 26.9% | 21.5% |
| VAT share (% of total revenue) *Eurostat, 1999-2019* | 16.6% | 19.1% | 18.7% | 18.1% |

Source: own calculations based on Eurostat and OECD data

A final difference in our SVAR model was the use of VAT revenues as a proxy for government revenues. Previous research suggested that the heavier reliance on nondistortionary taxes such as the VAT could be associated with faster growth and more efficient fiscal stimulus. The two countries that have had the highest VAT revenue share from the total government revenues between 1999 and 2019 (Hungary and Poland, see Table 2) have the highest multipliers. A government spending shock of one standard deviation is followed by an approximately 0.1-0.2% increase in revenues in Hungary and an approximately 0.3-0.4% increase in Poland. The Czech and Slovak revenues barely change. The catch here is that while Hungary has by far the highest VAT share from total revenues (19.6% compared to 18.8% in Poland), the Hungarian reaction to a spending shock is only half or third of the Polish reaction.

As indicated in the literature review section, other factors that were found to have an impact on the efficiency of a fiscal stimulus include fiscal space (how indebted the country is, and how expensive it is to finance the stimulus), trade openness, and the possibility of relying on foreign capital. Table 2 features some key indicators that can measure these factors. The indicators were calculated as the mean of the yearly values. We compare the means featured in Table 2 to the multipliers calculated from the IRFs because there are no reliable data available on a quarterly basis about government debt, import share, FDI stock, and net external debt, so these factors could not be included in our SVAR model. The reviewed studies have come to contradictory conclusions about the size of government debt: a larger debt could both increase and decrease the efficiency of a spending shock. Lower interest rates should make fiscal policy more efficient, as should a lower reliance on imports, as well as a higher reliance on external capital (credit and investments).

By comparing the indicators in Table 2 of the four countries with the estimated spending multipliers, we can check how our results compare to previous findings.

- Hungary and Poland have the two highest average debt/GDP ratios, which seems to indicate that higher debt goes together with higher spending multipliers. The fact that Hungary has been by far the most indebted country in the V4 makes this connection weaker, since the Hungarian multiplier is lower than the Polish one (a similar observation was made when we discussed the role of VAT revenues).
- Hungary and Poland have the two highest average interest rate levels (on 10-year government bonds), yet their economies react the most intensively to fiscal stimulus. Table 2 shows nominal rates, but Hungary and Poland have higher average rates than Czechia and Slovakia even if we adjust for inflation. This result goes against theoretical predictions and empirical findings and could partially be explained by foreign lending.
- Poland has the lowest import/GDP ratio and the highest spending multiplier, which supports the expectation that a fiscal stimulus is more successful in a country that relies less on imports. Hungary blurs the picture somewhat yet again with a high openness indicator and multiplier pairing.
- Table 2 includes two indicators that measure reliance on foreign capital: FDI stocks (measuring investments) and net external debt (measuring lending). FDI stock figures do not show any clear pattern, but in regard to foreign borrowing, Hungary and Poland again have the highest ratio. These two countries borrow the most from foreigners, which could help fiscal stimulus because it can reduce the crowding-out effect of government spending on investments.

The major limitation of this analysis is the simplicity of the SVAR model: other than the short-term interest rate, it does not include proxies that could show the direct impact of factors such as indebtedness, openness, or foreign borrowing. Another limitation is that we do not distinguish between recession and expansion periods.

Conclusion

This study investigates the impact of fiscal policy by developing an SVAR model and estimating IRFs and spending multipliers for Czechia, Hungary, Poland, and Slovakia. The results suggest that fiscal expansion has a larger than usual impact in the V4 countries (except for Slovakia): the estimated long-term (5-year) cumulative spending multipliers are 0.81 for Czechia, 1.14 for Hungary, 1.76 for Poland, and -0.18 for Slovakia (the Slovakian multiplier is not significant). Peak values are even slightly higher: 1.4 in Czechia (Q2), 1.42 in Hungary (Q1), 1.88 in Poland (Q2), and 0.4 in Slovakia (Q2).

The second finding of this study is that the heavier use of VAT-type nondistortionary taxes could increase the impact of a spending stimulus. VAT revenues were used as a proxy for total government revenues, and we found that a government spending shock of one standard deviation is followed by an approximately 0.1-0.2% increase in revenues in Hungary and an approximately 0.3-0.4% increase in Poland. Hungary and Poland earn the highest share of government revenues from VAT among the V4 countries, and the model generates the two highest multipliers for these two countries. The fact that the VAT ratio is highest in Hungary and the multiplier is highest in Poland makes this finding less conclusive.

Finally, this study discusses the possible influence of factors such as indebtedness, openness, and foreign borrowing. Hungary and Poland have the highest debt/GDP ratio and the highest external debt ratio. This suggests that more debt and more foreign debt coincide with higher spending multipliers. Poland, on the other hand, has by far the lowest import/GDP ratio in the region, which indicates that lower levels of openness yield higher spending multipliers.

Our results have limitations that are mentioned at the end of the discussion section, but if accepted, the following policy implications may be derived from them:

1. Fiscal stimulus, especially in Poland and Hungary, can be a powerful tool to tackle the problem of the recent economic slowdown in the region.

2. Stimulus is powerful even in a high-debt environment, especially if foreign borrowing reduces the crowding-out effect of government expansion. Poland and Hungary have higher spending multipliers despite having higher 10-year government bond yields. Our approach cannot tell, however, how high interest rates can climb before hurting the efficiency of fiscal policy.

3. More reliance on VAT revenues could increase the value of the spending multiplier.